\documentclass[a4paper,11pt]{article}
\usepackage{pos}

\usepackage{xcolor}
\usepackage{multirow}
\usepackage{capt-of}
\usepackage{siunitx}
\usepackage{graphicx}
\usepackage{slashed}
\usepackage{tabularx}
\usepackage{multirow}
\usepackage{braket}
\usepackage{amsfonts}
\usepackage{mathtools}
\usepackage{mathrsfs}
\usepackage{bbold}
\usepackage{comment}
\usepackage[compat=1.1.0]{tikz-feynman}
\usepackage{subcaption}

\pgfmathsetmacro\sizedot{1.1}
\pgfmathsetmacro\sizesqdot{1.5}
\pgfmathsetmacro\sizecrodot{1.0}

\newcommand{\coeff}[2]{ \mathcal{C}_{#1} ^{#2} } 
\newcommand{\op}[2]{ \mathcal{O}_{#1} ^{#2} } 
\newcommand{\adm}{ \Gamma}
\newcommand{\gl}[2]{ T_{#1} ^{#2}}
\newcommand{\order}[1] {\mathcal{O}\left( #1 \right)}

\newcommand{\highscale}{\Lambda}

\newcommand{\opdim}{\mathfrak{D}}

\newcommand{\lag}{ \mathcal{L}} %Lagrangian

\title{ Renormalization group running effects in
	$pp \to t\bar{t}h$ in the SMEFT}
\author*{Stefano Di Noi}
\affiliation{Dipartimento di Fisica e Astronomia “G. Galilei”, Universit\`a di Padova,\\
  Via Marzolo 8,  Padua, Italy}

\affiliation{Istituto Nazionale di Fisica Nucleare, Sezione di Padova,\\
  Via Marzolo 8, Padua, Italy}

\emailAdd{stefano.dinoi@phd.unipd.it}
\emailAdd{ramona.groeber@pd.infn.it}

\abstract{
We analyze the effects of renormalization group running of the Wilson coefficients in the SMEFT in the context of single Higgs production in association with a top-antitop pair. In particular, we analyze the differential cross section with respect to the Higgs trasnverse momentum.
We extend the usual $\order{\alpha_s}$ analysis including also the top Yukawa running effects, whose impact can be significant when large Wilson coefficients are considered.
We employ a dynamical and a fixed renormalization scale with different set-ups for the Wilson coefficients, defined at the TeV scale. Additionally, we comment on the accuracy of the widely-used first leading-logarithm approximation.
}

\FullConference{12th Large Hadron Collider Physics Conference (LHCP2024)\\
3-7 June, 2024\\
 Boston, \\USA}

\begin{document}
\maketitle
\section{Introduction}
The Standard Model of particle physics (SM) stands as one of the most important scientific achievements of the recent time. Despite that, several observations and theoretical puzzles suggest that it should be extended. 
So far, no clear evidence for New Physics (NP) has been observed. Additionally, it will not be possible to explore higher energy regions in the near future, diminuishing the chances of direct observation of new degrees of freedom. 

Effective Field Theories (EFTs) represent a general and efficient framework to parametrize small deviations from the SM, allowing for global analyses combining data from different experiments.  
Here, we employ the Standard Model Effective Field Theory (SMEFT), 
which extends the SM Lagrangian with a series of higher-dimensional operators $\op{i}{}$. The aforementioned operators contain SM fields only and must be invariant under the unbroken SM gauge group, namely $\mathbf{SU(3)}_{\mathrm{C}} \otimes\mathbf{SU(2)}_{\mathrm{L}}\otimes\mathbf{U(1)}_{\mathrm{Y}}$. We have: 
\begin{equation}
	\label{eq:SMEFTlag}
	\lag_{\textrm{SMEFT}} =\lag_{\textrm{SM}} + \sum_{\opdim_i>4} \frac{\coeff{i}{}}{\highscale^{\opdim_i-4}} \op{i}{},
\end{equation}
where  $\opdim_i$ denotes the dimension of $\op{i}{}$ in units of energy. The Wilson coefficients $\coeff{i}{}$ in Eq.~\eqref{eq:SMEFTlag} capture the impact of heavy BSM particles having masses $\Lambda$, with $\Lambda \gg v = 246\, \mathrm{GeV}$.
Under the assumption of conservation of baryon and lepton number, the first term in the expansion has $\opdim = 6$. A complete and non-redundant basis at this order is the so-called \textit{Warsaw basis} \cite{Grzadkowski:2010es}, consisting in 2499 independent operators in the most flavor-agnostic scenario.  
Already at this order, a plethora of new interactions arise \cite{Dedes:2017zog}, changing significantly the SM phenomenology.

\section{Running effects}

The precision obtained by current experimental measurements often requires higher order computations involving loop diagrams, which typically diverge. 
The renormalization procedure provides a strategy to eliminate the divergencies from on-shell matrix elements, providing finite and well-defined physical predictions. 
As a consequence, it induces an energy-scale dependence in the parameters of the theory, described by a system of coupled differential equations which go under the name of Renormalization Group Equations (RGEs). At dimension-six, the RGEs must be linear, allowing to write them as

\begin{equation}		
	\mu \frac{ d \coeff{i}{}( \mu)}{d \mu} = \frac{1}{16 \pi^2} { \color{black}{\adm_{ij} ( \mu)} \coeff{j}{}( \mu) },
\end{equation}
where we introduced $\Gamma$, known as Anomalous Dimension Matrix (ADM). The one-loop results are fully known \cite{rge1,rge2,rge3} and some partial results at two-loop level are available in \cite{Bern:2019wie, Bern:2020ikv, Jenkins:2023bls,DiNoi:2023ygk,DiNoi:2024ajj,Born:2024mgz}.

The energy-scale dependence of the
 ADM is fully contained in the couplings, allowing to decompose it as
$\adm_{ij} (\mu)$  $=g_1^2 (\mu) \adm_{ij} ^{(g_1^2)}$ $+ g_2^2(\mu) \adm_{ij} ^{(g_2^2)} + \ldots,$
where the matrices $ \adm_{ij} ^{(g_i^2)}$ are constant. Neglecting all the terms apart from the strong coupling provides an exactly solvable system, 
%\begin{equation}
%\adm_{ij} (\mu)= g_s^2 (\mu) { \color{black} \adm_{ij} ^{(g_s^2)}} ,	
%\end{equation}
which represent the most used approach in the inclusion of running effects \cite{Aoude:2022aro, Maltoni:2024dpn, Grazzini:2018eyk, Battaglia:2021nys,Maltoni:2016yxb, Heinrich:2024rtg}.

If additional interactions are considered, an analytic solution is not possible. The RGEs must be solved either numerically or by employing an approximation, for example via the first leading-logarithm solution:
\begin{equation} \label{eq:approximate}
	{\color{black}	\coeff{i}{}(\mu_{\textrm{F}}) } = {\color{black} \coeff{i}{}(\mu_{\textrm{I}}) }+ 
{\color{black}\adm_{ij}(\mu_{\textrm{I}}) \coeff{j}{}(\mu_{\textrm{I}})}  \frac{\log{\left( {\color{black}\mu_{\textrm{F}} }/ 	{\color{black}\mu_{\textrm{I}}}\right)}}{16 \pi^2}.
\end{equation}
This method is simple and fast, but it is reliable only if the energy scales $\mu_{\textrm{F}},\,\mu_{\textrm{I}}$ are close. 

\section{Implementation and results}
\begin{figure}[t]
	\centering
	\begin{subfigure}[t]{0.22\linewidth}
		\centering
		\begin{tikzpicture}[baseline=(a)]
			\begin{feynman}[small]
				\vertex (a) [violet,square dot,scale=\sizesqdot] {};
				\vertex  (gi2) [below left=40 pt of a]{\( g \)};
				\vertex  (gi1) [above left=40 pt of a]{\( g \)};
				
				\vertex (h) [below right = 40 ptof a] {\( h \)};
				\vertex (g) [dot, scale = \sizedot,above right = 16 pt of a] {};
				\vertex  (t1) [above right = of g]{\( t \)};
				\vertex  (t2) [below right = of g]{\( \bar{t} \)};
				\diagram* {
					(gi1)  -- [gluon] (a),
					(gi2)  -- [gluon] (a),
					(g)  -- [gluon] (a),
					(h) -- [scalar] (a),
					(t2) --[fermion] (g) --[fermion] (t1)
				};
			\end{feynman}
		\end{tikzpicture}
		\caption{}\label{fig:ggtthSMEFT1}
	\end{subfigure}
	\begin{subfigure}[t]{0.22\linewidth}
		\centering
		\begin{tikzpicture}[baseline=(a)]
			\begin{feynman}[small]
				\vertex (a) [orange,square dot,scale =\sizesqdot] {};
				\vertex  (gi2) [below left=35 pt of a]{\( g \)};
				\vertex  (gi1) [above left= 35 pt  of a]{\( g \)};
				\vertex (h) [below right = 35 pt  of a] {\( h \)};
				%				\vertex (g) [dot, scale = \sizedot,above right = 16 pt of a] {};
				\vertex  (t1) [above right = 35 pt  of a]{\( t \)};
				\vertex  (t2) [right = 27 pt   of a]{\( \bar{t} \)};
				\diagram* {
					(gi1)  -- [gluon] (a),
					(gi2)  -- [gluon] (a),
					(t2) --[fermion] (a) --[fermion] (t1),
					(h) --[scalar] (a)
				};
			\end{feynman}
		\end{tikzpicture}
		\caption{}\label{fig:ggtthSMEFT2}
	\end{subfigure}
	\begin{subfigure}[t]{0.22\linewidth}
		\centering
		\begin{tikzpicture}[baseline=(h)]
			\begin{feynman}[small]
				\vertex (q1) {\(  q \)};
				\vertex (qqg1) [dot, scale=\sizedot, below right=of q1] {};
				\vertex (q2) [below left =  of qqg1] {\( \bar{q} \)};
				\vertex (qqg2) [orange, square dot, scale=\sizesqdot,right = of qqg1] {};
				\vertex (h) [right = of qqg2] {\( h \)};                
				\vertex  (t1) [above right= of qqg2]{\(  t \)};
				\vertex  (t2) [below right= of qqg2]{\(  \bar{t} \)};
				\diagram* {
					(q1) --[fermion] (qqg1) -- [fermion] (q2),
					(qqg1) --[gluon] (qqg2),
					(t2) -- [fermion] (qqg2)  -- [fermion] (t1),
					(h) -- [scalar] (qqg2)
				};
			\end{feynman}
		\end{tikzpicture}
		\caption{}\label{fig:qqtthSMEFT1}
	\end{subfigure}
	\begin{subfigure}[t]{0.22\linewidth}
		\centering
		\begin{tikzpicture}[baseline=(h)]
			\begin{feynman}[small]
				\vertex (q1) {\(  q \)};
				\vertex (qqg1) [blue,square dot, scale=\sizesqdot, below right=36 pt of q1] {};
				\vertex (q2) [below left =36 pt  of qqg1] {\( \bar{q} \)};
				\vertex (h) [right= of qqg1] {\( h \)};                
				\vertex (tth) [dot, scale=\sizedot,below right = 14 pt of qqg1] {};
				\vertex  (t1) [above right=36 pt of qqg1]{\(  t \)};
				\vertex  (t2) [below right=36 pt of qqg1]{\(  \bar{t} \)};
				\diagram* {
					(q1) --[fermion] (qqg1) -- [fermion] (q2),
					(t2) --[fermion] (tth) -- [fermion] (qqg1)  -- [fermion] (t1),
					(h) -- [scalar] (tth)
				};
			\end{feynman}
		\end{tikzpicture}
		\caption{}\label{fig:qqtthSMEFT2}
	\end{subfigure}
	\caption{Tree-level diagrams for $pp\to t\bar{t}h$ in the SMEFT. The colored square dots denote insertions of SMEFT operators.} \label{fig:pptthSMEFT}
\end{figure}
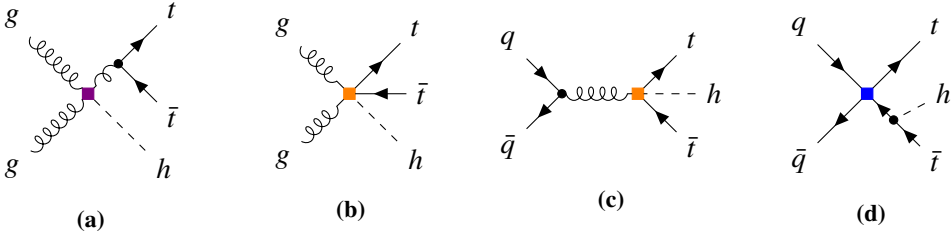

Higgs production in association with a top-antitop pair arises at tree-level in the SM, see Fig.~\ref{fig:pptthSM}. We consider only SMEFT operators which enter at tree-level, see Fig.~\ref{fig:pptthSMEFT}.

\begin{figure}[t]
	\centering
	\begin{subfigure}[t]{0.22\linewidth}
		\centering
		\begin{tikzpicture}[baseline=(h)]
			\begin{feynman}[small]
				\vertex  (gi1) {\(  g \)};
				\vertex (gi2) [below = of gi1] {\( g \)};
				\vertex (gtt1) [dot, scale = \sizedot, right = of gi1] {};
				\vertex (gtt2) [dot, scale = \sizedot, right = of gi2] {};
				\vertex (htt) [dot, scale=\sizedot, below =14pt of gtt1] {};
				\vertex (h) [right = of htt] {\( h \)};
				\vertex (tf1) [right = of gtt1] {\( t \)};
				\vertex (tf2) [right = of gtt2] {\( \bar{t} \)};
				\diagram* {
					(gi1)  -- [gluon] (gtt1),
					(gi2)  -- [gluon] (gtt2),
					(tf2)  -- [fermion] (gtt2) -- [fermion] (htt) -- [fermion] (gtt1) -- [fermion] (tf1),
					(h) -- [scalar] (htt)
				};
			\end{feynman}
		\end{tikzpicture}
		\caption{}\label{fig:ggtthSM1}
	\end{subfigure}
	\begin{subfigure}[t]{0.22\linewidth}
		\centering
		\begin{tikzpicture}[baseline=(h)]
			\begin{feynman}[small]
				\vertex  (gi1) {\(  g \)};
				\vertex (gi2) [below = of gi1] {\( g \)};
				\vertex (gtt1) [dot, scale = \sizedot, right = of gi1] {};
				\vertex (gtt2) [dot, scale = \sizedot, right = of gi2] {};
				\vertex (htt) [dot, scale=\sizedot, right=10pt of gtt1] {};
				\vertex (h) [below = 14 pt of tf1] {\( h \)};
				\vertex (tf1) [right = of gtt1] {\( t \)};
				\vertex (tf2) [right = of gtt2] {\( \bar{t} \)};
				\diagram* {
					(gi1)  -- [gluon] (gtt1),
					(gi2)  -- [gluon] (gtt2),
					(tf2)  -- [fermion] (gtt2)  -- [fermion] (gtt1) -- [fermion] (htt)-- [fermion] (tf1),
					(h) -- [scalar] (htt)
				};
		\end{feynman}  \end{tikzpicture}
		\caption{}\label{fig:ggtthSM2}
	\end{subfigure}
	\begin{subfigure}[t]{0.22\linewidth}
		\centering
		\begin{tikzpicture}[baseline=(h)]
			\begin{feynman}[small]
				\vertex (q1) {\(  q \)};
				\vertex (qqg1) [dot, scale=\sizedot, below right=of q1] {};
				\vertex (q2) [below left =  of qqg1] {\( \bar{q} \)};
				\vertex (qqg2) [dot, scale=\sizedot,right = of qqg1] {};
				\vertex (h) [right = of qqg2] {\( h \)};                
				\vertex (tth) [dot, scale=\sizedot,above right = 10 pt of qqg2] {};
				\vertex  (t1) [above right= of qqg2]{\(  t \)};
				\vertex  (t2) [below right= of qqg2]{\(  \bar{t} \)};
				\diagram* {
					(q1) --[fermion] (qqg1) -- [fermion] (q2),
					(qqg1) --[gluon] (qqg2),
					(t2) -- [fermion] (qqg2) --[fermion] (tth) -- [fermion] (t1),
					(h) -- [scalar] (tth)
				};
			\end{feynman}
		\end{tikzpicture}
		\caption{}\label{fig:qqtthSM1}
	\end{subfigure}
	\begin{subfigure}[t]{0.22\linewidth}
		\centering
		\begin{tikzpicture}[baseline=(h)]
			\begin{feynman}[small]
				\vertex (q1) {\(  q \)};
				\vertex (qqg1) [dot, scale=\sizedot, below right=of q1] {};
				\vertex (q2) [below left =  of qqg1] {\( \bar{q} \)};
				\vertex (qqg2) [dot, scale=\sizedot,right = of qqg1] {};
				\vertex (h) [right= of qqg2] {\( h \)};                
				\vertex (tth) [dot, scale=\sizedot,below right = 10 pt of qqg2] {};
				\vertex  (t1) [above right= of qqg2]{\(  t \)};
				\vertex  (t2) [below right= of qqg2]{\(  \bar{t} \)};
				\diagram* {
					(q1) --[fermion] (qqg1) -- [fermion] (q2),
					(qqg1) --[gluon] (qqg2),
					(t2) --[fermion] (tth) -- [fermion] (qqg2)  -- [fermion] (t1),
					(h) -- [scalar] (tth)
				};
			\end{feynman}
		\end{tikzpicture}
		\caption{}\label{fig:qqtthSM2}
	\end{subfigure}
	\caption{Tree-level diagrams for $pp\to t\bar{t}h$ in the SM.} \label{fig:pptthSM}
\end{figure}
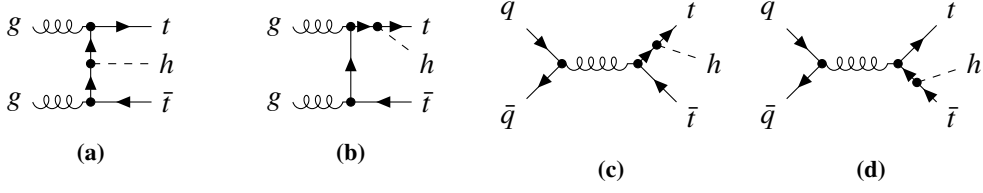

To assess the impact of the running effects, we set some non-vanishing Wilson coefficients at $\Lambda=2\; \text{TeV}$ and we run them to the renormalization scale $\mu_{\mathrm{R}}$ using \texttt{RGESolver} \cite{DiNoi:2022ejg}. 
We repeat the computation with two different renormalization scales: a fixed scale $\mu_{\mathrm{R}}=m_t$ (same for all the events) and a
dynamical scale $\mu_{\mathrm{R}}=(p_{T,h}+p_{T,t}+p_{T, \bar{t}})/2$ (differs event by event). 
Understanding if the usage of a fixed renormalization scale is a valid approximation is crucial, due to the technical difficulties associated to the numeric solution of the RGEs.

\begin{figure}[h!]
	\centering
	\begin{minipage}[b]{0.48\textwidth}
		\centering
		\includegraphics[width=\linewidth]{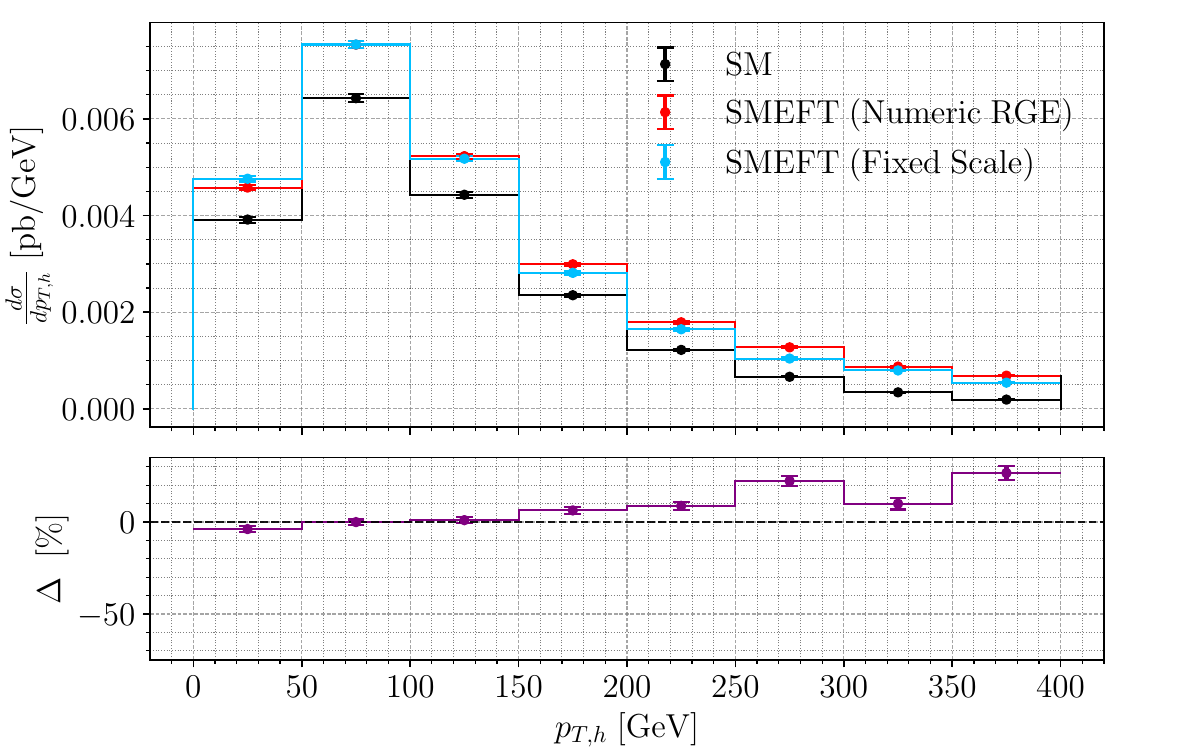}
		\caption{Higgs transverse momentum distribution in the conservative scenario, dynamical vs fixed renormalization scale.}
		\label{fig:dsigmaMaxL4}
	\end{minipage} \hfill
	\begin{minipage}[b]{0.48\textwidth}
		\centering
		\includegraphics[width=\linewidth]{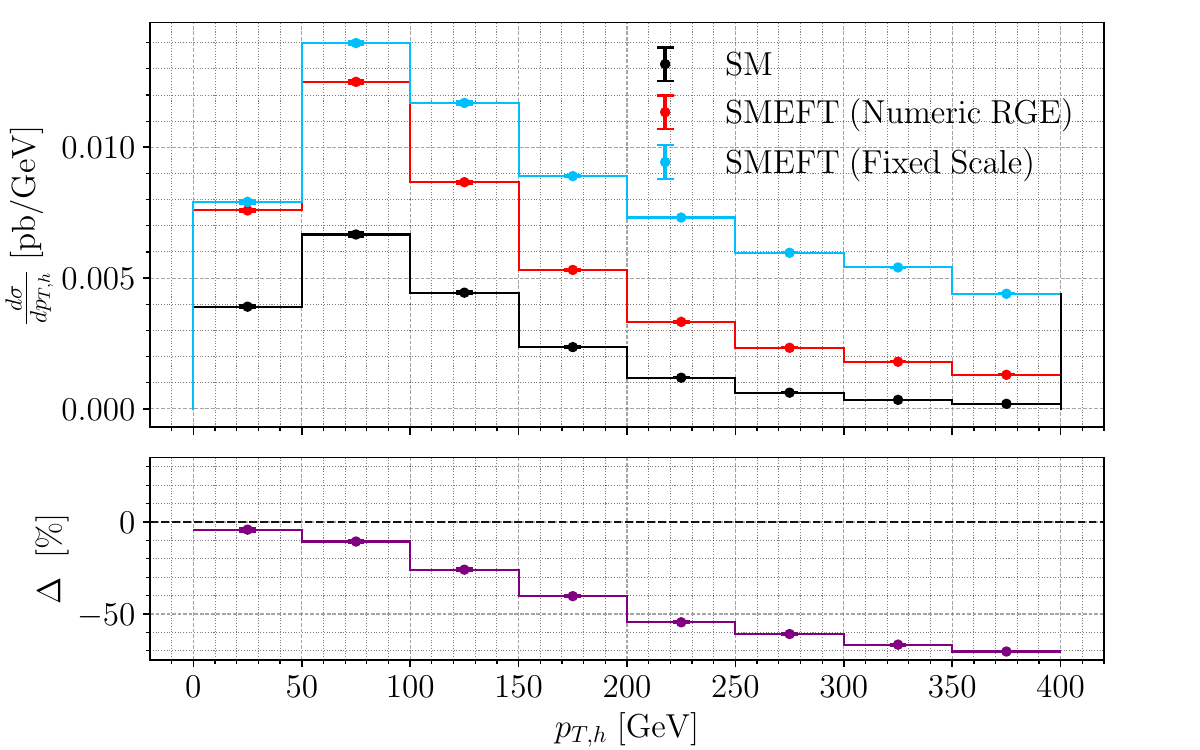}
		\caption{Higgs transverse momentum distribution in the extreme scenario, dynamical vs fixed renormalization scale.}
		\label{fig:dsigmaMaxL2}
	\end{minipage}
\end{figure}

We compare two scenarios: we employ the conservative $\order{1/\Lambda^4}$ bounds and the extreme $\order{1/\Lambda^2}$ bounds from \cite{Ethier:2021bye}, showing the results in Figs.~\ref{fig:dsigmaMaxL4},~\ref{fig:dsigmaMaxL2}. We observe a difference between the two renormalization scale choices up to 25\% in the first case, increasing to 70\% in the latter.

\begin{figure}[h!]
	\centering
	\begin{minipage}[b]{0.48\textwidth}
		\centering
		\includegraphics[width=\linewidth]{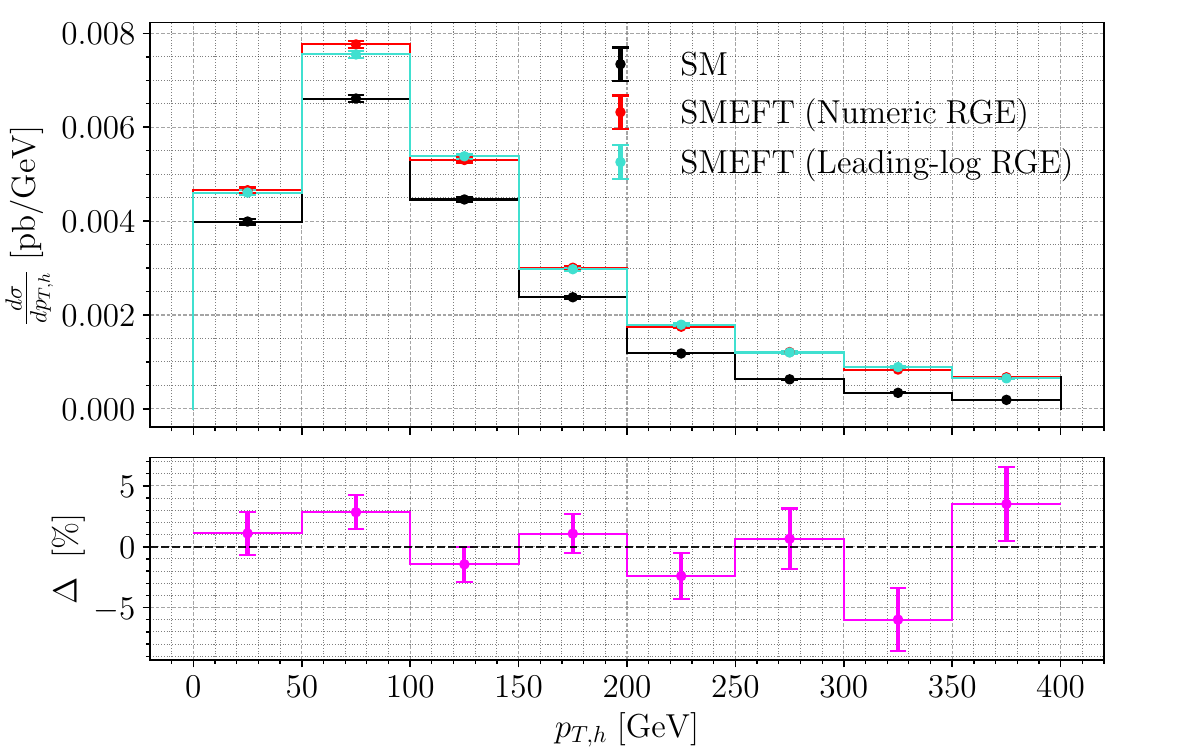}
			\caption{Higgs transverse momentum distribution in the conservative scenario, numeric vs leading-logarithm running.}
		\label{fig:dsigmaCompL4}
	\end{minipage} \hfill
	\begin{minipage}[b]{0.48\textwidth}
		\centering
		\includegraphics[width=\linewidth]{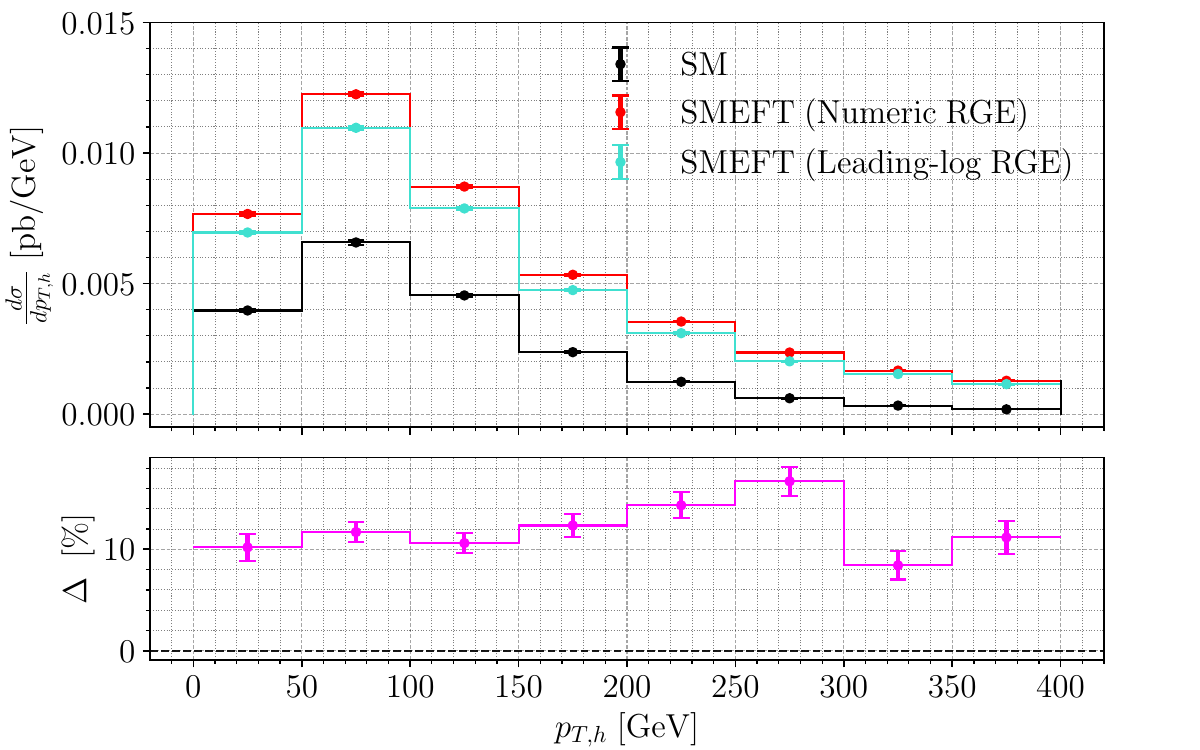}
	\caption{Higgs transverse momentum distribution in the extreme scenario, numeric vs leading-logarithm running.} \label{fig:dsigmaCompL2}
	\end{minipage}
\end{figure}

We have compared the numeric solution of the RGEs with the first leading-logarithm approximation in Eq.~\eqref{eq:approximate}, observing negliglible differences between the two in the conservative scenario in Fig.~\ref{fig:dsigmaCompL4}, with instead discrepancies up to 15\% in the extreme case in Fig.~\ref{fig:dsigmaCompL2}. 

\begin{figure}[h!]
	\centering
	\begin{minipage}[b]{0.48\textwidth}
		\centering
		\includegraphics[width=\linewidth]{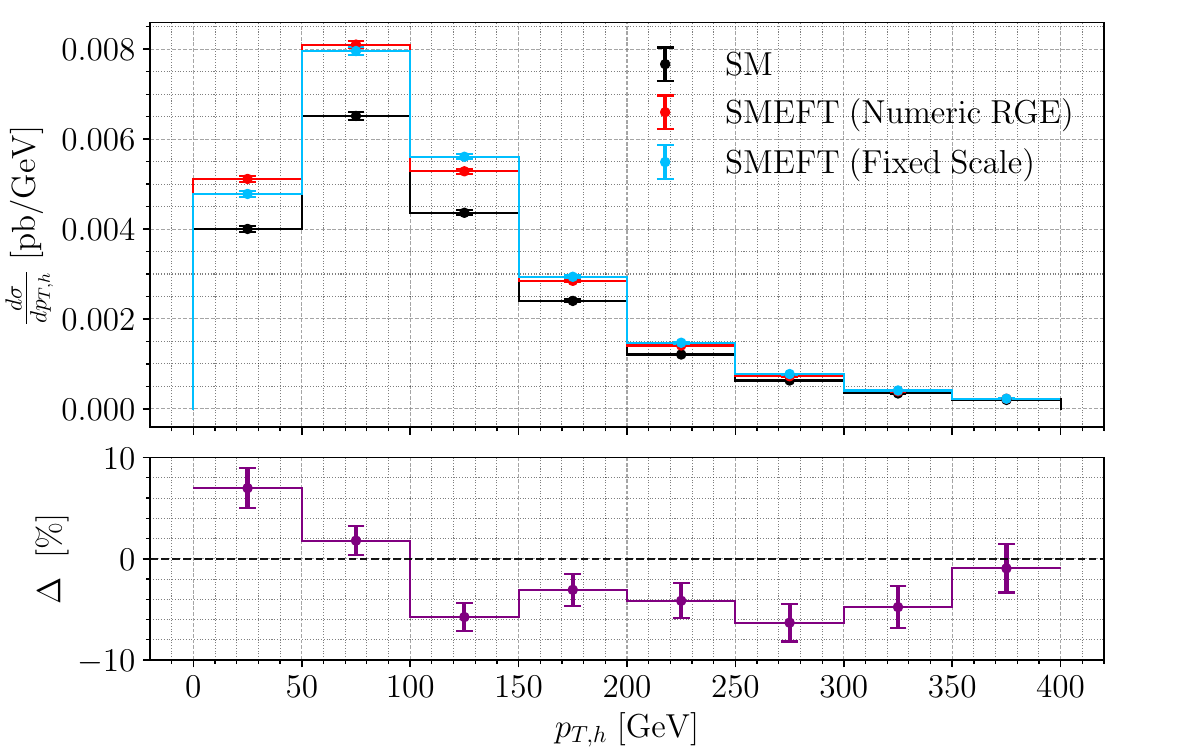}
		\caption{Higgs transverse momentum distribution with $\coeff{Qt}{(1)}(\Lambda) = \frac{4}{3} \times 20 \, / \mathrm{TeV}^2$.}
		\label{fig:dsigmaQt1}
		\includegraphics[width=\linewidth]{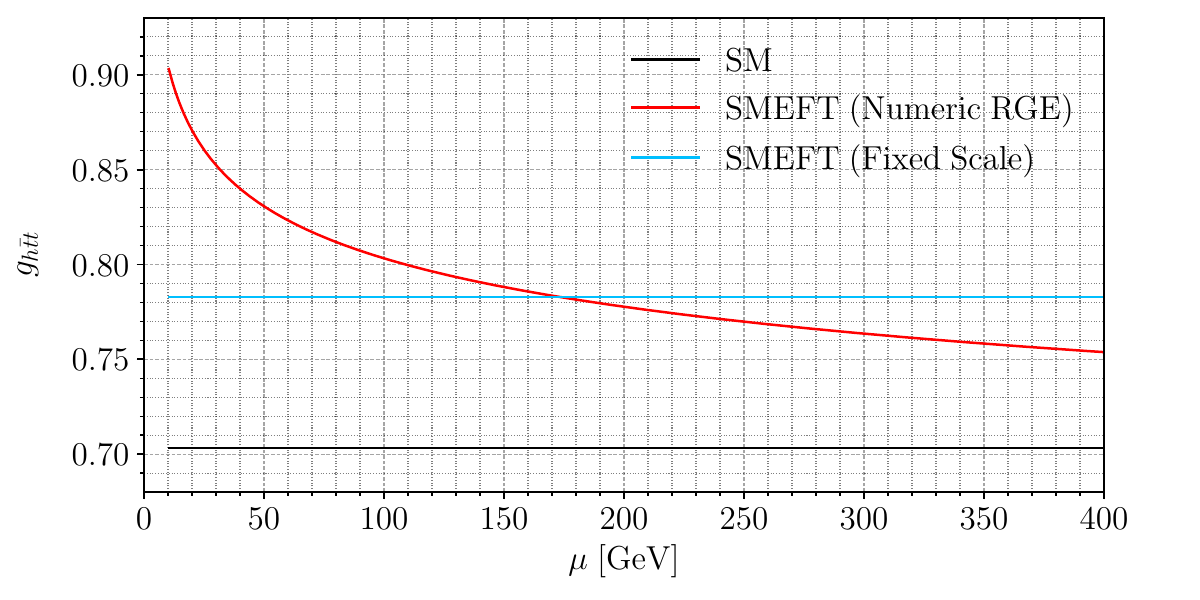}
		\caption{Running of $g_{h t\bar{t}}$ with $\coeff{Qt}{(1)}(\Lambda) = \frac{4}{3} \times 20 \, / \mathrm{TeV}^2$.}
		\label{fig:ghttRunningQt1}
	\end{minipage}\hfill
	\begin{minipage}[b]{0.48\textwidth}
		\centering
		\includegraphics[width=\linewidth]{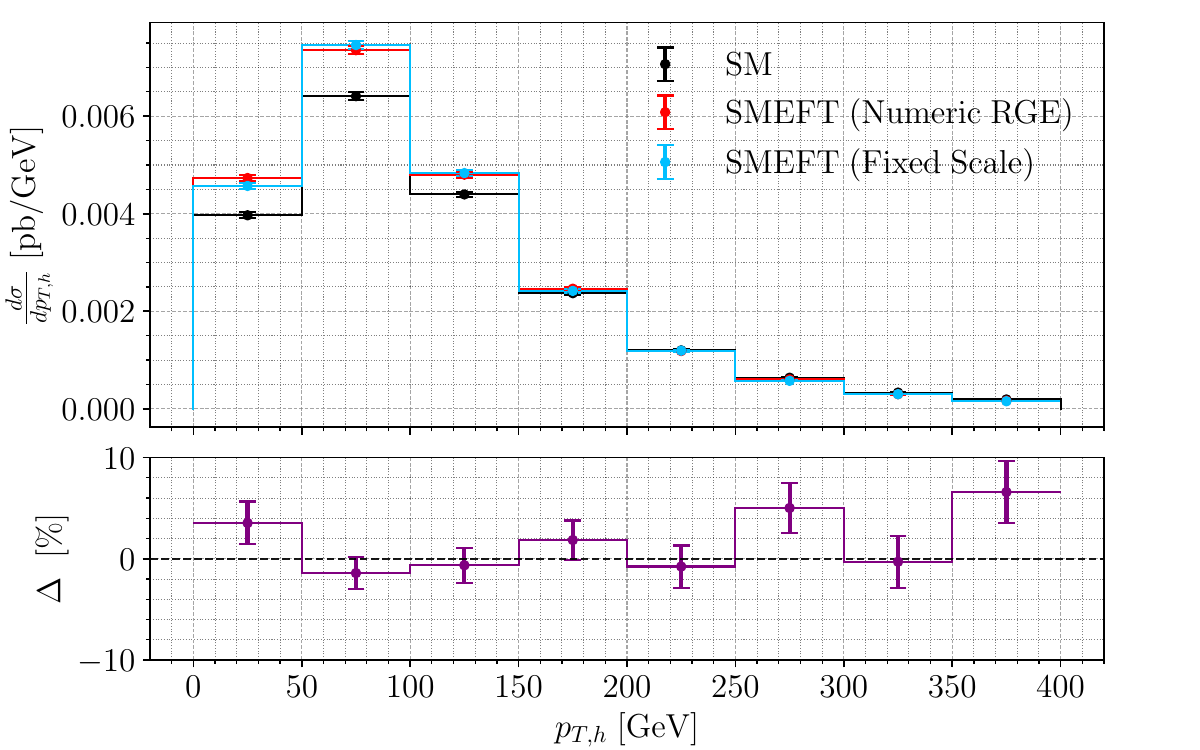}
		\caption{Higgs transverse momentum distribution with $\coeff{Qt}{(8)}(\Lambda) = 20 \, / \mathrm{TeV}^2$.}
		\label{fig:dsigmaQt8}
		\includegraphics[width=\linewidth]{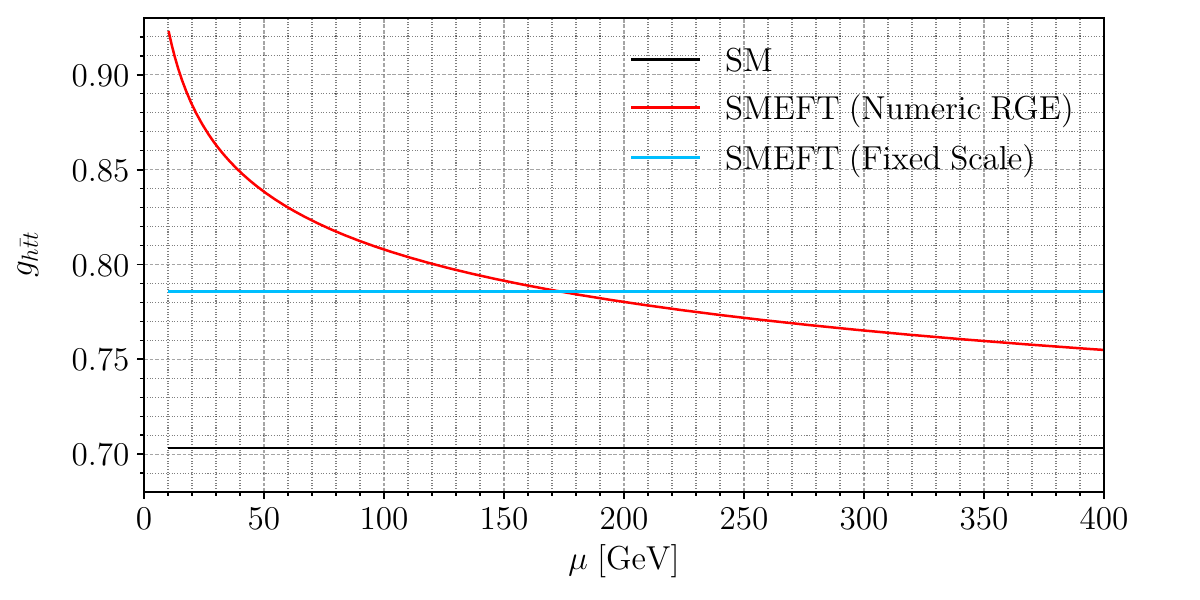}
		\caption{Running of $g_{h t\bar{t}}$ with $\coeff{Qt}{(8)}(\Lambda) = 20 \, / \mathrm{TeV}^2$.}
		\label{fig:ghttRunningQt8}
	\end{minipage}
\end{figure}

We studied the differential distribution with respect to the Higgs transverse momentum considering individually the two four-top operators  $\op{Qt}{(1,8)} = \left(\bar{Q}_L \gamma^\mu (\gl{}{A}) Q_L\right) \left(\bar{t}_R \gamma_\mu (\gl{}{A}) t_R \right)$ in Figs.~\ref{fig:dsigmaQt1},~\ref{fig:dsigmaQt8}. Their contribution to $g_{h t\bar{t}}$ , the effective Higgs-top coupling, goes as $y_t^3 \left(\coeff{Qt}{(1)}+ \frac{4}{3} \coeff{Qt}{(8)} \right)$, providing a similar behaviour within both scenarios, see Figs.~\ref{fig:ghttRunningQt1},~\ref{fig:ghttRunningQt8}. 
Instead, their strong mixing with other operators entering in $pp \to \bar{t}t h$ at tree-level is different, due to the different color structure.
By observing that there is a sizeable difference between dynamical and fixed renormalization scales in both cases, we conclude that top-Yukawa running effects can be relevant in presence of large Wilson coefficients, advocating for their inclusion together with effects proportional to the strong coupling.

\section{Conclusions and outlook}
Running effects are expected to become more and more relevant in the near future, due to the increasing precision at the experimental and theoretical level. 
We have discussed that relevant differences can arise when employing a dynamical renormalization scale
with respect to a fixed renormalization scale by analyzing the transverse momentum spectrum in single Higgs production in association with a top-antitop pair.
Additionally, we have observed that the widely-used leading-logarithm approximation in the solution of the RGEs departures sizeably from the numeric one in presence of large Wilson coefficients. 
Finally, we showed how Yukawa contributions can be as relevant as strong ones in some scenarios, highlighting the importance of their inclusion in phenomenological studies. 
\bibliographystyle{unsrt} % Or use \bibliographystyle{plain}

\begin{comment}

\end{comment}
\bibliographystyle{JHEP}
\bibliography{references}
\end{document}